\documentclass[thmsa]{article}
\usepackage{amssymb}

\usepackage{sw20lart}

\input{tcilatex}
\begin{document}

\title{Weak Transversality and Partially Invariant Solutions}
\author{A. M. Grundland\thanks{%
e-mail: grundlan@CRM.UMontreal.ca}, P. Tempesta\thanks{%
e-mail: tempesta@CRM.UMontreal.ca} and P. Winternitz\thanks{%
e-mail: wintern@CRM.UMontreal.ca} \\
Centre de Recherches Math\'{e}matiques, Universit\'{e} de Montr\'{e}al, \\
C. P. 6128, succ. Centre--ville, Montr\'{e}al, Qu\'{e}bec H3C 3J7, Canada}
\date{June 3, 2002}
\maketitle

\begin{abstract}
New exact solutions are obtained for several nonlinear physical equations,
namely the Navier--Stokes and Euler systems, an isentropic compressible
fluid system and a vector nonlinear Schr\"{o}dinger equation. The solution
methods make use of the symmetry group of the system in situations when the
standard Lie method of symmetry reduction is not applicable.
\end{abstract}

\section{Introduction}

Lie group theory provides very general and efficient methods for obtaining
exact analytic solutions of systems of partial differential equations,
specially nonlinear ones \cite{Olver}--\cite{Vino}. The different methods
have in common that they provide a reduction of the original system. This
reduction usually means the reduction of the number of independent variables
occurring, possibly a reduction of the number of dependent ones too.

The ''standard'', or ''classical'' reduction method goes back to the
original work of Sophus Lie and is explained in many modern texts \cite
{Olver}--\cite{Vino}. Essentially, it amounts to requiring that a solution
of the equation should be invariant under some subgroup $G_{0}\subseteq G$,
where $G$ is the symmetry group of the considered system of equations. The
subgroup $G_{0}\,$must satisfy certain criteria, in order to provide such
group invariant solutions (see below).

The purpose of this article is to further develop, compare and apply
alternative reduction methods. They have in common the fact that they
provide solutions not obtainable by Lie's classical method. We shall survey
the ''tool kit'' available for obtaining particular solutions of systems of
partial differential equations, and further refine some of the tools. In the
process we shall obtain new solutions of some physically important equations
such as the Navier--Stokes equations, the Euler equations, the equations
describing an isentropic compressible fluid model and the vector nonlinear
Schr\"{o}dinger equation.

We shall consider a system of $m$ partial differential equations of order $n$%
, involving $p$ independent variables $\left( x_{1},x_{2},...,x_{p}\right) $
and $q$ dependent variables $\left( u_{1},u_{2},...,u_{q}\right) $%
\begin{equation}
\Delta _{\nu }\left( x,u^{\left( n\right) }\right) =0,\qquad \nu =1,...,m,
\label{I.1}
\end{equation}
where $u^{\left( n\right) }$ denotes all partial derivatives of $u_{\alpha }$%
, up to order $n$.

S. Lie's classical method of symmetry reduction consists of several steps.

1. Find the local Lie group $G$ of local point transformations taking
solutions into solutions. Realize its Lie algebra $L$ in terms of vector
fields and identify it as an abstract Lie algebra. The vector fields will
have the form

\begin{equation}
\mathbf{v}_{a}\,\mathbf{=}\sum_{i=1}^{p}\xi _{a}^{i}\left( x,u\right) \frac{%
\partial }{\partial x^{i}}+\sum_{\alpha =1}^{q}\varphi _{a}^{\alpha }\left(
x,u\right) \frac{\partial }{\partial u^{\alpha }},\qquad a=1,..,r=\dim G.
\label{I.2}
\end{equation}

2. Classify the subalgebras $L_{i}\subset L\,$into conjugacy classes under
the action of the largest group $\widetilde{G}$ leaving the system of
equations invariant (we have $G\subseteq \widetilde{G}$). Choose a
representative of each class.

3. For each representative subalgebra $L_{i}\,$find the invariants $%
I_{\gamma }\left( x_{1},...,x_{p},u_{1},...,u_{q}\right) $ of the action of
the group $G_{i}=\left\langle \exp L_{i}\right\rangle $ in the space $%
M\thicksim X\times U$ of independent and dependent variables. Let us assume
that $p+q-s$ functionally independent invariants $I_{\gamma }$ exist, at
least locally ($s$ is the dimension of the generic orbits of $G_{i}$).

4. Divide (if possible) the invariants $I_{\gamma }\,$into two sets: $%
\left\{ F_{1},...,F_{q}\right\} $and $\left\{ \xi _{1},...,\xi _{k}\right\} $%
, $k=p-s$ in such a manner that the Jacobian relating $F_{\mu }$ and $%
u_{\alpha }$ $\left( \alpha ,\mu =1,...,q\right) $ $\,$has maximal rank 
\begin{equation}
J=\frac{\partial \left( F_{1},...,F_{q}\right) }{\partial \left(
u_{1},...,u_{q}\right) },\qquad rank\,J=q.  \label{I.3}
\end{equation}
Then consider $F_{\mu }\,$ to be functions of the other invariants $\left\{
\xi _{i}\right\} $which serve as the new independent variables and use
condition (\ref{I.3}) to express the dependent variables $u_{\alpha }$ in
terms of the invariants.

5. Substitute the obtained expressions for $u_{\alpha }$ into the original
system and obtain the reduced system, involving only invariants. The reduced
system will involve only $k=p-s$, $s\geq 1$ independent variables.

6. Solve the reduced system. If the variables $\xi ^{i}$ depend on $%
x_{1},...,x_{p}$ only, this will yield explicit solutions of system (\ref
{I.1}). Otherwise, if $\xi _{i}$ depend also on the original dependent
variables $u_{\alpha }$, we obtain implicit solutions.

7. Apply a general symmetry group transformation to these solutions.

This procedure provides a family of particular solutions that can be used to
satisfy particular boundary or initial conditions. The classification of
subgroups $G_{i}$ can be viewed as a classification of conditions that can
be imposed on the obtained solutions.

Steps 3, 4 and 5 of Lie's method can be reformulated as follows. Take the
vector fields $\left\{ \mathbf{v}_{a}\right\} $ forming a basis of the
considered Lie subalgebra $\frak{g}_{i}\subset \frak{g}$ and set their
characteristics equal to zero 
\begin{equation}
Q_{a}^{\alpha }=\varphi _{a}^{\alpha }\left( x,u\right) -\sum_{i=1}^{p}\xi
_{a}^{i}\left( x,u\right) \,u_{x_{i}}^{\alpha }=0,\quad \alpha
=1,...,q,\quad a=1,...,r_{i}=\dim \frak{g}_{i}.  \label{I.4}
\end{equation}
Solve the systems (\ref{I.1}) and (\ref{I.4}) simultaneously.

Several alternative reduction procedures have been proposed, going beyond
Lie's classical method and providing further solutions. They all have in
common that they add some system of equations to the original system (\ref
{I.1}) and that the extended system is solved simultaneously. These
additional equations replace the characteristic system (\ref{I.4}).

In its generality, this was proposed as the method of ''differential
constraints'' \cite{SSY}, and independently as the method of ''side
conditions'' \cite{Olver-Ros1, Olver-Ros2}. Different methods differ by the
choice of this system of side conditions.

Basically two different approaches exist in the literature. The first makes
further use of the symmetry group $G$ of system (\ref{I.1}), the second
approach goes beyond this group of point transformations, or even gives up
group theory altogether. Let us briefly discuss some of these methods.

1. ''Group invariant solutions without transversality''. This method was
proposed by Anderson et al. \cite{AFT} quite recently and deals with the
situation when the rank condition (\ref{I.3}) is not satisfied. It was shown 
\cite{AFT} that under certain conditions on the subgroup $G_{i}\subset G$
one can still obtain $G_{i}$ invariant solutions.We recall that a group
invariant solution, with or without transversality, is transformed into
itself by the subgroup $G_{i}\subset G$.

2. The method of partially invariant solutions. A solution $u\left( x\right) 
$ is ''partially invariant'' \cite{Ovs, Ovs2} under a subgroup $G_{i}\subset
G\,$ of the invariance group, if $G_{i}$, when acting on $u\left( x\right)
\, $sweeps out a manifold of a dimension that is larger than that of the
graph of the solution, but less than the dimension of the entire space $M$.
A group $G_{i}$ may provide both invariant and partially invariant solutions
(see below). However, if the rank condition (\ref{I.3}) for the Jacobian is
not satisfied, at least on a set of solutions, then $G_{i}$ will not provide
invariant solutions, but may provide partially invariant ones. If the rank
of the Jacobian $J\,$ is $q^{\prime }$, with $q^{\prime }<q$, then we can
express $u_{1},...,u_{q^{\prime }}$ in terms of invariants and let $%
u_{q^{\prime }+1},...,u_{q}$ depend on all of the original variables $%
x_{1},...,x_{p}.$ We then substitute the dependent variables back into the
original system and obtain a ''partial reduction''. Solving this system, we
obtain the partially invariant solutions. Irreducible partially invariant
solutions are partially invariant solutions that cannot be obtained by Lie's
method using the subgroup $G_{i}\,$or any other subgroup $G_{i}^{\prime }$
of the symmetry group of the considered equations. Such solutions were
constructed for certain nonlinear Klein--Gordon and Schrodinger equations,
in \cite{MW} and \cite{MSW}, and for some equations of hydrodynamics in Ref. 
\cite{GL1, GL2, GL3}.

The theory of partially invariant solutions was further developed by Ondich 
\cite{Ondich1, Ondich2} who formulated irreducibility criteria for certain
classes of equations. For other applications, see Ref. \cite{Ibrag2, Ibrag3,
SDR}.

In terms of the equations (\ref{I.4}), the method described above boils down
to taking only $q^{\prime }<q$ of the equations (\ref{I.4}). As we shall
show, it is possible to obtain further partially invariant solutions by
different methods.

Among methods that go beyond the use of the symmetry group $G$ we mention
the following.

3. The Clarkson--Kruskal direct reduction method \cite{Clark-Krus, CM} does
not make explicit use of group theory. It is postulated that the dependent
variables should be expressed in terms of new dependent variables that
depend on fewer independent ones. The corresponding Ansatz is substituted
into the original equation, which must then be solved. It has been shown
that the direct method is intimately related to the method of conditional
symmetries \cite{CW, Pucci, Levi-Wint}, to the ''nonclassical method''
proposed by Bluman and Cole \cite{Blum-Cole, ABH}, and to potential
symmetries \cite{Sacco}. This method can also be interpreted in terms of
side conditions \cite{Olver-Ros1, Olver-Ros2}. The differential constraints
added to system (\ref{I.1}) in this case have the form of first order
quasilinear partial differential equations of the form (\ref{I.4}). However
the coefficients $\varphi _{a}^{\alpha }$ and $\xi _{\alpha }^{i}$ are not
related to a Lie point symmetry of eq. (\ref{I.1}).

4. The group foliation method. The method goes back to S. Lie, is described
by Ovsiannikov \cite{Ovs} and has recently been applied to obtain solutions
of self dual Einstein equations \cite{NS, MShW}. In terms of differential
constraints this method amounts to embedding the system (\ref{I.1}) into a
larger system, consisting of all equations up to some definite order,
invariant under the same Lie point symmetry group as (\ref{I.1}) (and
involving the same variables).

5. The method of ''partial Lie--point symmetries'', proposed by Cicogna and
Gaeta \cite{CG}. This is a modification of the method of conditional
symmetries. The method is in some cases easier to use and may provide
solutions in cases when the equations of the conditional symmetry method
prove to be untractable.

6. For integrable equations \cite{AC} generalized symmetries can be used to
generate side conditions. These will be higher order equations, rather than
first order ones.

7. The method of nonlocal symmetries. This consists in extending the space
of dependent variables by adding some auxiliary variables, which can be
potentials or pseudopotentials associated to the system of equations under
analysis \cite{BRK, Vino, KV, LLST}. A nonlocal symmetry will be then a
symmetry of the original system augmented with the equations defining the
new nonlocal variables.

This article is organized as follows. In Section 2 we introduce the concept
of ''weak transversality'' allowing us to simplify the method of ''group
invariant solutions without transversality'' \cite{AFT} and to relate the
rank of the matrix of invariants (\ref{I.3}) to that of the coefficients of
vector fields. The method of weak transversality is applied in Section 3 to
obtain new invariant solutions of the Navier--Stokes equations and of the
isentropic compressible fluid model. In Section 4 we establish a relation
between partially invariant solutions and the transversality condition. This
is then applied to obtain new irreducible partially invariant solutions of
the vector nonlinear Schr\"{o}dinger equation, of the Euler equations, of
the Navier--Stokes equations and of the isentropic compressible fluid model
in (3+1) dimensions. The concept of the irreducibility of partially
invariant solutions is discussed. Some conclusions are presented in the
final Section 5.

\section{Group invariant solutions, strong and weak transversality}

S. Lie's classical method of symmetry reduction was outlined in the
Introduction. The first two steps are entirely algorithmic and we shall
assume that they have already been performed. Thus, we are given a system of
equations (\ref{I.1}) and have found its Lie point symmetry group $G$, the
Lie algebra of which is the symmetry algebra $L$. The symmetry algebra has
dimension $r$ and has a basis realized by vector fields of the form (\ref
{I.2}). Each vector field has the property that its $n$--th prolongation
annihilates the system (\ref{I.1}) on its solution set 
\begin{equation}
pr\,\mathbf{v}\,\Delta _{\nu }\mid _{\Delta _{\mu }=0}=0,\qquad \nu ,\,\mu
=1,...,m.  \label{5}
\end{equation}
The functions $\xi _{a}^{i}\left( x,u\right) \,$and $\varphi _{a}^{i}\left(
x,u\right) $ are thus explicitly known.

Let us now consider a subgroup $G_{0}\subset G\,$and its Lie algebra $\frak{g%
}_{0}$. A solution $u=f\left( x\right) $ of the system (\ref{I.1}) is $G_{0}$
invariant if its graph $\Gamma _{f}\sim \left\{ x,f\left( x\right) \right\}
\,$is a $G_{0}$ invariant set: 
\begin{equation}
g\cdot \Gamma _{f}=\Gamma _{f},\qquad g\in G_{0}.  \label{6}
\end{equation}
The vector field (\ref{I.2}) can be written in evolutionary form \cite{Olver}
as 
\begin{equation}
\mathbf{v}_{E,a}=\sum_{\alpha =1}^{q}\left( \varphi _{a}^{\alpha }\left(
x,u\right) -\sum_{i=1}^{p}\xi _{a}^{i}\left( x,u\right) u_{x_{i}}^{\alpha
}\right) \partial _{u_{\alpha }},\quad a=1,...,r.  \label{7}
\end{equation}

A $G_{0}$--invariant solution will satisfy the $q\times $ $r_{0}$
characteristic equations (\ref{I.4}) associated with the basis elements of
the Lie algebra $L_{0}\,\left( \dim L_{0}=r_{0}\right) $.

The following matrices play an essential role in symmetry reduction using
the symmetry group of the considered system of equations.

1. The matrices $\Xi _{1}$ and $\Xi _{2}$ of the coefficients of the vector
fields $\mathbf{v}_{a}$ spanning the algebra $L$, or its subalgebra $L_{0}$,
and defined as follows: 
\[
\Xi _{1}=\left\{ \xi _{a}^{i}\left( x,u\right) \right\} ,\qquad \Xi _{1}\in 
\Bbb{R}^{r\times p} 
\]
\begin{equation}
\Xi _{2}=\left\{ \xi _{a}^{i}\left( x,u\right) ,\varphi _{a}^{\alpha }\left(
x,u\right) \right\} ,\qquad \Xi _{2}\in \Bbb{R}^{r\times \left( p+q\right) }
\label{8}
\end{equation}
where $a=1,...,r$ labels the rows, $i\,$and $\alpha \,$labels the columns.

2. The matrix of characteristics of the vector fields (\ref{I.2}) (or (\ref
{7})) spanning the considered algebra $L\,$(or its subalgebra $L_{0}$) 
\begin{equation}
Q_{a}^{\alpha }=\left\{ \mathbf{v}_{a}^{E}u_{\alpha }\right\} \qquad
a=1,...,r;\quad \,\alpha =1,...,q.  \label{9}
\end{equation}

3. The Jacobian matrix $J$ of the transformation relating the dependent
variables $u_{\alpha }$ and the invariants of the action of $G_{0}$ on the
space $M\sim X\times U$ of independent and dependent variables.

Let us now consider a specific subalgebra $L_{0}\subset L$ and use it to
obtain group invariant solutions via symmetry reduction. If the group $%
G_{0}\,$acts regularly and transversally on $M\sim X\times U$ then 
\begin{equation}
rank\left\{ \xi _{a}^{i}\left( x,u\right) \right\} =rank\left\{ \xi
_{a}^{i}\left( x,u\right) ,\varphi _{a}^{\alpha }\left( x,u\right) \right\} .
\label{10}
\end{equation}

This rank is equal to the dimension of the generic orbits of $G_{0}$ on $M$.
If the transversality condition (\ref{10}) is satisfied, at least locally,
for all $\left\{ x,u\right\} \in M$, then Lie's classical reduction method
is directly applicable. Indeed, if (\ref{10}) is satisfied then the rank of
the matrix $J$ of eq. (\ref{I.3}) is maximal, $rank\,J=q$ (for a proof, see
e.g. Ref. \cite{Olver}, Chapter 3.5). It follows that all dependent
variables can be expressed in terms of invariants and a reduction is
immediate (to a system with $q$ dependent variables and $p-s$ independent
ones).

If the action of $G_{0}$ on $M\,$is fiber preserving (i.e. the new
independent variables only depend on the old independent ones), Lie's method
provides explicit solutions. This happens because the new invariant
independent variables $z_{i}$ can be chosen to depend only on the original
independent variables 
\begin{equation}
z_{i}=z_{i}\left( x_{1},...,x_{p}\right) \qquad i=1,...,p-s.  \label{11}
\end{equation}
More generally, if we have $z_{i}=z_{i}\left( x,u\right) $, we obtain
implicit solutions.

We shall call the rank condition (\ref{10}) ''strong transversality''. Quite
recently \cite{AFT} a method was proposed for obtaining group invariant
solutions when equation (\ref{10}) is not satisfied. The method of Ref. \cite
{AFT} can actually be simplified by introducing the concept of ''weak
transversality''.

\begin{definition}
The local transversality condition will be said to be satisfied in the weak
sense if it holds only on a subset $\widetilde{M}$ $\subset M,$ rather than
on the entire space $M$: 
\begin{equation}
rank\left\{ \xi _{a}^{i}\left( x,u\right) \right\} \mid _{\widetilde{M}%
}=rank\left\{ \xi _{a}^{i}\left( x,u\right) ,\varphi _{a}^{\alpha }\left(
x,u\right) \right\} \mid _{\widetilde{M}}  \label{WT}
\end{equation}
\end{definition}

In other words, even if the transversality condition is not in general
satisfied, there may exist a class $S$ of functions $u=f\left( x\right) $
such that for them the condition (\ref{WT}) holds.

The ''weak transversality'' method is quite simple, when applicable. It
consists of several steps.

1. Determine the conditions on the functions $u=f\left( x\right) \,$under
which eq. (\ref{WT}) is satisfied. Solve these conditions to obtain the
general form of these functions.

2. Substitute the obtained expressions into the matrix of characteristics (%
\ref{9}) and require that the condition $rank\,Q=0$ be satisfied. This
further constrains the functions $f\left( x\right) $.

3. Substitute the obtained expressions into the system (\ref{I.1}). By
construction, the solutions, if they exist, will be $G_{0}$--invariant.

This method can only be applied if the matrix elements in the matrix $\Xi
_{2}$ depend explicitly on the variables $u_{\alpha }$. This poses strong
restrictions on the considered algebra $\frak{g}_{0}$. We shall give some
examples of the method in Section 3.

\section{Examples of invariant solutions obtained by the weak transversality
method.}

\subsection{The Navier--Stokes equations}

The Navier--Stokes equations in (3+1) dimensions describing the flow of an
incompressible viscuous fluid are:

\begin{equation}
\overrightarrow{u_{t}}+\overrightarrow{u}\cdot \nabla \overrightarrow{u}%
+\nabla p-\nu \nabla ^{2}\overrightarrow{u}=0,  \label{NS1}
\end{equation}
\begin{equation}
\nabla \cdot \overrightarrow{u}=0,  \label{NS2}
\end{equation}
where $\overrightarrow{u}=\left( u_{1}\left( x,y,z,t\right) ,u_{2}\left(
x,y,z,t\right) ,u_{3}\left( x,y,z,t\right) \right) $ is the velocity field, $%
p=p\left( x,y,z,t\right) $ the fluid pressure and $\nu \,$ the viscosity
coefficient.

The symmetry properties of these equations have been intensively
investigated by many authors from different points of view (see, for
instance, \cite{GLST} and references therein). It is well known \cite{Ll}
that eqs. (\ref{NS1})--(\ref{NS2}) are invariant under the flow generated by
the following vector fields: 
\begin{equation}
B_{1}=\alpha \,\partial _{x}+\stackrel{\cdot }{\alpha }\partial _{u_{1}}-%
\stackrel{\cdot \cdot }{\alpha }x\,\partial _{p},  \label{NS3}
\end{equation}
\begin{equation}
B_{2}=\beta \,\partial _{y}+\stackrel{\cdot }{\beta }\partial _{u_{2}}-%
\stackrel{\cdot \cdot }{\beta }y\,\partial _{p},  \label{NS4}
\end{equation}
\begin{equation}
B_{3}=\gamma \,\partial _{z}+\stackrel{\cdot }{\gamma }\partial _{u_{3}}-%
\stackrel{\cdot \cdot }{\gamma }z\,\partial _{p},  \label{NS5}
\end{equation}
\begin{equation}
T=\partial _{t},  \label{NS6}
\end{equation}
\begin{equation}
Q=\partial _{p},  \label{NS7}
\end{equation}
\begin{equation}
D=x\partial _{x}+y\partial _{y}+z\partial _{z}+2t\partial _{t}-u_{1}\partial
_{u_{1}}-u_{2}\partial _{u_{2}}-u_{3}\partial _{u_{3}}-2p\partial _{p},
\label{NS8}
\end{equation}
\begin{equation}
L_{1}=z\partial _{y}-y\partial _{z}+u_{3}\partial _{u_{2}}-u_{2}\partial
_{u_{3},}  \label{NS9}
\end{equation}
\begin{equation}
L_{2}=x\partial _{z}-z\partial _{x}+u_{1}\partial _{u_{3}}-u_{3}\partial
_{u_{1}},  \label{NS10}
\end{equation}
\begin{equation}
L_{3}=y\partial _{x}-x\partial _{y}+u_{2}\partial _{u_{1}}-u_{1}\partial
_{u_{2}},  \label{NS11}
\end{equation}
where $\alpha ,\beta ,\gamma $ and $\delta $ are arbitrary functions of
time. The operators $B_{j}\,$generate symmetry transformations that can be
interpreted as boosts to frames moving with arbitrary velocities $%
\overrightarrow{v\left( t\right) }=\lambda \left( \stackrel{\cdot }{\alpha },%
\stackrel{\cdot }{\beta ,}\stackrel{\cdot }{\gamma }\right) $, where $%
\lambda $ is a constant$.$ Space translations and Galilei boosts are
obtained if $\alpha ,\beta $ and $\gamma $ are linear in $t$ .The operators $%
T$ and $Q$ express the invariance of the eqs. (\ref{NS1})--(\ref{NS2}) under
translations of time and pressure, $D$ generates scaling transformations,
and $L_{1},L_{2}$ and $L_{3}$ are the generators of the group of the
rotations of the Euclidean space.

\textbf{Example 1. }Let us consider the subalgebra generated by $L_{1},L_{2}$
and $L_{3}$\textit{. }Here we apply the ideas discussed in Section 2 to
determine rotationally invariant solutions for the Navier--Stokes equations.

The matrices of the coefficients $\Xi _{1}=\left( \xi _{a}^{i}\left(
x,u\right) \right) $ and $\Xi _{2}=\left( \xi _{a}^{i}\left( x,u\right)
,\phi _{a}^{\alpha }\left( x,u\right) \right) $ are represented by 
\begin{equation}
\Xi _{1}=\left( 
\begin{array}{lll}
0 & z & -y \\ 
-z & 0 & x \\ 
y & -x & 0
\end{array}
\right) ,\qquad \qquad \Xi _{2}=\left( 
\begin{array}{llllll}
0 & z & -y & 0 & u_{3} & -u_{2} \\ 
-z & 0 & x & -u_{3} & 0 & u_{1} \\ 
y & -x & 0 & u_{2} & -u_{1} & 0
\end{array}
\right)
\end{equation}

We observe that the matrix $\Xi _{1}$ has rank 2, whereas the matrix $\Xi
_{2}$ has rank 3. In this situation, the transversality condition is
violated in the strong sense. In other words, it is not true that for 
\textit{every\thinspace }$\,$function $\overrightarrow{u}=\left(
u_{1},u_{2},u_{3}\right) $ the ranks of the matrices $\Xi _{1}$ and $\Xi
_{2} $ coincide. In general, the system of characteristics is not
compatible, the Jacobian matrix $J$ will not have maximal rank, and it is 
\textit{not} possible to use the classical symmetry reduction approach. To
overcome these difficulties, let us force the matrix $\Xi _{2}$ $\,$to be of
rank 2. This requirement is equivalent to a system of algebraic equations
for $\overrightarrow{u}$, obtained imposing that the determinants of all $%
3\times 3\,$matrices constructed using the rows and the columns of $\Xi _{2}$
be equal to zero. Once this algebraic system is solved, we get the class $S$
of functions $\overrightarrow{u}=\left( u_{1},u_{2},u_{3}\right) $ on which
transversality is weakly restored. The class $S$ in this case is
characterized by the conditions: 
\begin{equation}
u_{1}=f\left( x,y,z,t\right) \,x,\quad u_{2}=f\left( x,y,z,t\right)
\,y,\quad u_{3}=f\left( x,y,z,t\right) \,z,\quad p=p\left( x,y,z,t\right)
\label{Sl1}
\end{equation}
The second step consists in solving the characteristic system $Q_{k}^{\alpha
}\left( x,u^{\left( 1\right) }\right) =0$ for the class $S$ of eqs. (\ref
{Sl1}). This forces the functions $f$ and $p$ to have the form:

\begin{equation}
f=f\left( r,t\right) ,\qquad p=p\left( r,t\right) ,  \label{fp}
\end{equation}
where $r=\sqrt{x^{2}+y^{2}+z^{2}}.$ Relations (\ref{fp}) represent the most
general form for the function $\overrightarrow{u}$ and $p$ to be
rotationally invariant in the Euclidean space. Substituting these
expressions for $\overrightarrow{u}$ and $p$ into the eqs. (\ref{NS1})--(\ref
{NS2}), we find the solution 
\begin{equation}
\overrightarrow{u}=\frac{a\left( t\right) }{r^{3}}\overrightarrow{x},\quad p=%
\frac{\stackrel{\cdot }{a}}{r}-\frac{a^{2}}{2r^{4}}+b.  \label{sol}
\end{equation}

The same vector fields $L_{1},L_{2}$ and $L_{3}$ provide also a subalgebra
of the symmetry algebra of the Euler equations. Anderson et al.\cite{AFT}
obtained a class of rotationally invariant solutions of the Euler equations
by means of their technique of reduction diagrams. In Ref. \cite{GL1}
partially invariant solutions related to the subalgebra $\left\{
L_{1},L_{2},L_{3}\right\} $ for the equations describing a nonstationary and
isentropic flow for an ideal and compressible fluid in $\left( 3+1\right) $
dimensions have been constructed using the transformation (\ref{Sl1}). A
similar situation is also observed in magnetohydrodynamics \cite{GL2}. Our
solutions of the Navier--Stokes equations coincide with the solutions of the
Euler equations found in Ref. \cite{AFT}. Physically that means that the
solutions (\ref{sol}) describe a laminar flow for which viscosity plays no
role. This phenomenon occours because the components of the vector $%
\overrightarrow{u}$ in eq. (\ref{sol}) are all harmonic functions, i.e. they
satisfy the Laplace equation, in addition to the Navier--Stokes equations.

\textbf{Example 2}. Now, Let us analyze the subalgebra defined by the
operators 
\[
D=x\partial _{x}+y\partial _{y}+z\partial _{z}+2t\partial _{t}-u_{1}\partial
_{u_{1}}-u_{2}\partial _{u_{2}}-u_{3}\partial _{u_{3}}-2p\partial _{p}, 
\]
\[
L_{3}=y\partial _{x}-x\partial _{y}+u_{2}\partial _{u_{1}}-u_{1}\partial
_{u_{2}}, 
\]
\[
X=t^{k}\,\partial _{x}+k\,t^{k-1}\partial _{u_{1}}-k\,\left( k-1\right)
\,t^{k-2}x\,\partial _{p}, 
\]
\begin{equation}
Y=t^{k}\,\partial _{y}+k\,t^{k-1}\partial _{u_{2}}-k\,\left( k-1\right)
\,t^{k-2}y\,\partial _{p},  \label{g2}
\end{equation}
which is a subalgebra of the Galilei--similitude algebra for a given $k\in 
\Bbb{R}$. It is immediate to check that the local transversality is violated
for this subalgebra in the strong sense. The matrix $\Xi _{2}$ is now
represented by 
\[
\left( 
\begin{array}{llllllll}
x & y & z & 2t & -u_{1} & -u_{2} & -u_{3} & -2p \\ 
y & -x & 0 & 0 & u_{2} & -u_{1} & 0 & 0 \\ 
t^{k} & 0 & 0 & 0 & k\,t^{k-1} & 0 & 0 & -k\,\left( k-1\right) \,t^{k-2}x \\ 
0 & t^{k} & 0 & 0 & 0 & k\,t^{k-1} & 0 & -k\,\left( k-1\right) \,t^{k-2}y
\end{array}
\right) 
\]
If we impose that the matrix $\Xi _{2}$ should have rank $3$ (that is weak
transversality), we get 
\[
u_{1}=k\frac{x}{t},\qquad u_{2}=k\frac{y}{t} 
\]
\[
u_{3}=u_{3}\left( x,y,z,t\right) ,\qquad p=p(x,y,z,t). 
\]

As a second step, let us solve the characteristic system $Q_{a}^{\alpha
}\left( x,u^{\left( 1\right) }\right) =0,$ which consists of 16 linear
differential equations of first order in the derivatives of the velocity
components $u_{j}$ and the pressure $p$. The most general function living in
the space of the dependent variables and invariant under the flow associated
to the generators (\ref{g2}) is 
\begin{equation}
u_{1}=k\frac{x}{t},\quad u_{2}=k\frac{y}{t},\quad u_{3}=\frac{\alpha
\QOVERD( ) {t}{z^{2}}}{z^{2}}  \label{S21}
\end{equation}
\begin{equation}
p=-\frac{k\left( k-1\right) \left( x^{2}+y^{2}\right) }{2t^{2}}+\frac{\beta
\QOVERD( ) {t}{z^{2}}}{z^{2}},  \label{S24}
\end{equation}
with $\alpha $ and $\beta \,$arbitrary functions of $t$ $z^{-2}$.
Substituting into the Navier--Stokes equations (\ref{NS1})--(\ref{NS2}), we
obtain the following three parameter set of solutions: 
\begin{equation}
u_{1}=k\frac{x}{t},\quad u_{2}=k\frac{y}{t},\quad u_{3}=\frac{c_{1}}{\sqrt{t}%
}-2\frac{kz}{t}  \label{S25}
\end{equation}
\begin{equation}
p=\frac{1}{2t^{2}}\left\{ c_{1}\sqrt{t}z+k\left( x^{2}+y^{2}+4c_{1}\sqrt{t}%
z-2z^{2}\right) -k^{2}\left( x^{2}+y^{2}+4z^{2}\right) +2c_{2}t\right\} ,
\label{S26}
\end{equation}
with $c_{1},c_{2},k\in \Bbb{R}$. This solution is invariant under the group
generated by the algebra (\ref{g2}) and satisfies weak (but not strong)
transversality.

\subsection{The isentropic compressible fluid model}

The equations describing the non--stationary isentropic flow of a
compressible ideal fluid are \cite{Ovs3} 
\begin{equation}
\overrightarrow{u_{t}}+\overrightarrow{u}\cdot \nabla \overrightarrow{u}%
+k\,a\,\nabla a=0  \label{IF1}
\end{equation}
\begin{equation}
a_{t}+\overrightarrow{u}\cdot \nabla a+k^{-1}a\,\nabla \cdot \overrightarrow{%
u}=0,  \label{IF2}
\end{equation}
where $\overrightarrow{u}=u_{1}\left( x,y,z,t\right) ,\,u_{2}\left(
x,y,z,t\right) ,\,u_{3}\left( x,y,z,t\right) $ is the velocity field, $%
a=a\left( x,y,z,t\right) $ is the velocity of sound, related to the pressure 
$p$ and the density $\rho $ by the formula $a=\left( \frac{\gamma \,p}{\rho }%
\right) ^{1/2}$, $\gamma $ is the adiabatic exponent and $k=2/\left( \gamma
-1\right) $. The symmetry group $G$ of eqs. (\ref{IF1})--(\ref{IF2}) was
derived in ref. \cite{GL3}. For $k\neq 3$, $G$ it is generated by the
following vector fields: 
\begin{equation}
P_{0}=\partial _{t},\quad P_{1}=\partial _{x},\quad P_{2}=\partial
_{y},\quad P_{3}=\partial _{z}  \label{IFA1}
\end{equation}
\begin{equation}
K_{1}=t\partial _{x}+\partial _{u_{1}},\quad K_{2}=t\partial _{y}+\partial
_{u_{2}},\quad K_{3}=t\partial _{z}+\partial _{u_{3}}
\end{equation}
\begin{equation}
L_{1}=z\partial _{y}-y\partial _{z}+u_{3}\partial _{u_{2}}-u_{2}\partial
_{u_{3},}
\end{equation}
\begin{equation}
L_{2}=x\partial _{z}-z\partial _{x}+u_{1}\partial _{u_{3}}-u_{3}\partial
_{u_{1}},
\end{equation}
\begin{equation}
L_{3}=y\partial _{x}-x\partial _{y}+u_{2}\partial _{u_{1}}-u_{1}\partial
_{u_{2}},
\end{equation}
\begin{equation}
F=x\partial _{x}+y\partial _{y}+z\partial _{z}+t\partial _{t},\quad
G=-t\partial _{t}+u_{1}\partial _{u_{1}}+u_{2}\partial
_{u_{2}}+u_{3}\partial _{u_{3}}+a\,\partial _{a}.  \label{IFA6}
\end{equation}

We mention that for $k=3$ the symmetry algebra contains an additional
element generating projective transformations. The operators $P_{i}\,,K_{i}$
and $L_{i}$ are the infinitesimal generators of space translations, Galilei
boosts and rotations, respectively. The operators $F$ and $G$ generate
scaling transformations.

\smallskip \textbf{Example 3. }Let us consider the subalgebra $\left\{
L_{3},F+G,K_{1},K_{2}\right\} .$ The matrix $\Xi _{2}$ is given by 
\begin{equation}
\left( 
\begin{array}{llllllll}
y & -x & 0 & 0 & u_{2} & -u_{1} & 0 & 0 \\ 
x & y & z & 0 & u_{1} & u_{2} & u_{3} & a \\ 
t & 0 & 0 & 0 & 1 & 0 & 0 & 0 \\ 
0 & t & 0 & 0 & 0 & 1 & 0 & 0
\end{array}
\right)  \label{IF3}
\end{equation}
and the transversality is again violated in the strong sense, because $%
rank\,\Xi _{1}=3$ and $rank\,\Xi _{2}=4.$ If we force the matrix $\Xi _{2}$
to be of rank 3, then we get the following constraints 
\begin{equation}
u_{1}=\frac{x}{t},\quad u_{2}=\frac{y}{t},\quad u_{3}=u_{3}\left(
x,y,z,t\right) ,\quad a=a\left( x,y,z,t\right) .  \label{IF4}
\end{equation}

\smallskip From the characteristic system $Q_{a}^{\alpha }\left( x,u^{\left(
1\right) }\right) =0$ we deduce 
\begin{equation}
u_{3}=z\,W\left( t\right) ,\qquad a=z\,A\left( t\right) ,  \label{IF5}
\end{equation}
where $W$ and $A$ are arbitrary functions of time. Now, substituting
relations (\ref{IF4})--(\ref{IF5}) into the system (\ref{IF1})--(\ref{IF2})
we obtain the relation 
\begin{equation}
A\left( t\right) =\sqrt{-\frac{1}{k}\left( W^{2}+W^{\prime }\right) }
\label{IF6}
\end{equation}
and a second order ODE for $W$%
\begin{equation}
W^{\prime \prime }+2\left( 2+\frac{1}{k}\right) WW^{\prime }+2\left( 1+\frac{%
1}{k}\right) W^{3}+\frac{4}{k\,t}\left( W^{\prime }+W^{2}\right) =0
\label{IF7}
\end{equation}

In general, eq. (\ref{IF7}) does not have the Painlev\'{e} property. For
special values of the parameter $k$, namely $k=-1\,$and $k=-2$, it does. In
these cases it can be reduced to a canonical form (see Ref. \cite{Ince},
p.334) via a linear transformation of the type 
\begin{equation}
W=\alpha \left( t\right) \,U\left( z\left( t\right) \right) +\beta \left(
t\right) .  \label{IF8}
\end{equation}

1. For $k=-1$ we have 
\begin{equation}
W^{\prime \prime }=-2WW^{\prime }+p\left( t\right) \,\left( W^{\prime
}+W^{2}\right) ,  \label{IF9}
\end{equation}
where $p\left( t\right) =4/t$. This equation can be integrated and its
solution, which is regular, is 
\begin{equation}
W=\frac{c_{1}t^{2}\left( I_{-\frac{5}{6}}\left( \frac{c_{1}t^{3}}{3}\right)
+c_{2}I_{\frac{5}{6}}\left( \frac{c_{1}t^{3}}{3}\right) \right) }{I_{\frac{1%
}{6}}\left( \frac{c_{1}t^{3}}{3}\right) +c_{2}I_{-\frac{1}{6}}\left( \frac{%
c_{1}t^{3}}{3}\right) },
\end{equation}
where $c_{1}$ and $c_{2}$ are constants and $I_{n}\left( x\right) $ is the
modified Bessel function of the first kind. Correspondingly we find 
\begin{equation}
A=c_{1}t^{2}
\end{equation}
in eq. (\ref{IF6}).

2. For $k=-2$ we have 
\begin{equation}
W^{\prime \prime }=-3WW^{\prime }-W^{3}+q\left( t\right) \left( W^{\prime
}+W^{2}\right) ,
\end{equation}
where $q\left( t\right) =2/t$. In this case, we can integrate and the
general solution is: 
\begin{equation}
W=\frac{4\,t^{3}+c_{1}}{t^{4}+c_{1}\,t+c_{2}}.
\end{equation}
The solution of eq. (\ref{IF6}) is

\smallskip 
\begin{equation}
A=2\sqrt{3}\sqrt{\frac{t^{2}}{t^{4}+c_{1}t+c_{2}}},
\end{equation}
where $c_{1}$, and $c_{2}$ are constants. The solutions for $k=-1$ and $k=-2$
represent nonscattering waves.

\section{Partially invariant solutions of systems of differential equations
and the transversality condition}

A useful tool applicable to the study of systems of differential equations,
and intimately related to the standard Lie approach, is the theory of
partially invariant solutions. The relevant notion in this context is the
defect $\delta \,$of a $k$--dimensional manifold $M$ with respect to a Lie
group $G$. When the group acts on a $p$--dimensional submanifold $\Gamma
\subseteq M$, it sweeps out an orbit $G\left( \Gamma \right) $. The manifold 
$\Gamma $ will be identified with the graph $\Gamma _{f}$ of a function $%
u=f\left( x\right) $, so its dimension will coincide with the number of
independent variables (also denoted $p$). As we already said, the case $%
G\left( \Gamma _{f}\right) =\Gamma _{f}\,$ corresponds to the $G$%
--invariance of the manifold. Otherwise, $G\left( \Gamma _{f}\right) \,$will
be a more generic subset of $M$. There is no guarantee that this subset will
be a submanifold. However, if the intersection between an orbit $O$ of $G$
and $\,\Gamma _{f}$ has a dimension which is constant in a neighbourhood $N$
of a point of $\Gamma _{f}$, then there exists a neighbourhood $\widetilde{G}
$ of the identity of $G\,\,$such that the subset $\widetilde{G}$ $\left(
N\cap \Gamma _{f}\right) $ is a submanifold \cite{Ondich1}. In the
subsequent considerations, $G\left( \Gamma _{f}\right) \,$will be considered
as a submanifold.

Let $G$ be a group, acting regularly with $s$--dimensional orbits. We call
the number 
\begin{equation}
\delta =\dim G\left( \Gamma _{f}\right) -\dim \Gamma _{f}  \label{1.12}
\end{equation}
\newline
the defect $\delta $ of the function $f$ with respect to $G$ . The usual $G$%
--invariant functions correspond to the case $\delta =0$. A function will be
said to be \textit{generic} if $\delta =m_{0}=\min \left\{ s,k-p\right\} .\,$%
The more interesting situation is when $0<\delta <m_{0}$, which is the case
we will dealing with. In this case, the function $f$ will be said to be
partially invariant \cite{Ovs}.

Let us consider the system (\ref{I.1}) of partial differential equations,
whose symmetry group $G$ acts on the $p+q$--dimensional space $M=X\times U$.
Let $\frak{g}$ be a subalgebra of the symmetry algebra of $\Delta $, and $Q$
the characteristic matrix associated to the set of its generators. Then $%
u=f\left( x\right) $ is a partially invariant solution of $\Delta $ with
defect $\delta $ with respect to $\frak{g}$ if and only if \cite{Ovs,
Ondich1, Olver2} 
\begin{equation}
rank\left( Q\left( x,u^{\left( 1\right) }\right) \right) =\delta .
\label{1.14}
\end{equation}

The condition (\ref{1.14}) provides a system of differential equations
involving the dependent variables $u=\left( u_{1},...u_{q}\right) $. In
order to determine partially invariant solutions, we can extend the original
system $\Delta $ by adding the set of differential constraints given by the
condition (\ref{1.14}). We must then solve the extended system consisting of
eq. (\ref{I.1}) and (\ref{1.14}). The set of equations given by the
prescription (\ref{1.14})\thinspace is less constraining than the set
required to obtain $G$--invariance, as in formulas (\ref{I.4}).

In this section we will study the role of the local transversality condition
(and in particular of the notion of weak transversality) in the theory of
partially invariant solutions and propose a strategy to find them.

Let us start by noticing that a partially invariant solution of a system of
differential equations can be naturally related to the violation of the
transversality condition. Indeed, let $\Delta \left( x,u^{\left( n\right)
}\right) =0$ be a system of differential equations defined over $M\subset
X\times U$ and $u_{0}=u_{0}\left( x_{0}\right) \,$be a solution of $\Delta $%
. Let $G$ be an $r$--dimensional subgroup of the symmetry group of $\Delta $%
, acting regularly on $M$, whose generators are given by (\ref{I.2}). If the
condition 
\begin{equation}
rank\left( \xi _{a}^{i}\left( x_{0},u_{0}\right) \right) <rank\left( \xi
_{a}^{i}\left( x_{0},u_{0}\right) ,\phi _{a}^{\alpha }\left(
x_{0},u_{0}\right) \right)  \label{TC}
\end{equation}
is satisfied, then $u_{0}=f\left( x_{0}\right) $ $\,$is a partially
invariant solution of $\Delta $ (or possibly a generic one).

\textbf{Example 4. }The vector nonlinear Schr\"{o}dinger equation 
\begin{equation}
i\psi _{t}+\Delta \psi =\left( \overline{\psi }\psi \right) \psi ,
\label{VNSE}
\end{equation}

where $\psi \in \Bbb{C}^{N}$ and $\Delta $ is the Laplace operator in $n$
dimensions, plays an important role in many areas of physics. For instance,
in nonlinear optics it describes the interaction of electromagnetic waves
propagating with different polarizations in nonlinear media \cite{BZ}. In
hydrodynamics, it furnishes a model for the description of the interactions
of $N$ water waves in a deep fluid \cite{Benney}--\cite{Phillips}. For these
and other applications of the vector nonlinear Schr\"{o}dinger equation, see
also Ref. \cite{Rem}.

Let us consider the case of three components $\left( N=3\right) $ and two
spatial dimensions $\left( n=2\right) $. The symmetry algebra has been
computed in Ref. \cite{SW}. In terms of amplitude and phase, the components
of the wave function will be written as $\psi _{i}=\rho _{i}e^{\omega _{i}}.$
In particular, we will discuss the role of the subalgebra 
\begin{equation}
\left\{ \partial _{x},\,\partial _{y},\,y\,\partial _{x}-x\,\partial
_{y}+a_{1}\partial \omega _{1}+a_{2}\partial \omega _{2}+a_{3}\partial
\omega _{3}\right\} ,  \label{subSE}
\end{equation}
generated by the two translations in the plane and a rotation combined with
a transformation of the phases. The matrix of the coefficients $\Xi _{2}$,
given by 
\[
\left( 
\begin{array}{llllllll}
1 & 0 & 0 & 0 & 0 & 0 & 0 & 0 \\ 
0 & 1 & 0 & 0 & 0 & 0 & 0 & 0 \\ 
y & -x & a_{1} & a_{2} & a_{3} & 0 & 0 & 0
\end{array}
\right) 
\]
has rank 3, unless $\left( a_{1},a_{2},a_{3}\right) =\left( 0,0,0\right) $.
Let us consider the case $a_{2}=a_{3}=0,$ $a_{1}\neq 0$. We obtain $\rho
_{i}=\rho _{i}\left( t\right) $, $\left( i=1,2,3\right) $, $\omega
_{1}=\omega _{1}\left( t\right) ,\omega _{2}=\omega _{2}\left( t\right) $,
but $\omega _{3}$ is not an invariant, so we keep $\omega _{3}=\omega
_{3}\left( x,y,z,t\right) $. Substituting into eq. (\ref{VNSE}), we obtain: 
\begin{equation}
\rho _{1}=\frac{\gamma _{1}}{\sqrt{t\left( t-t_{0}\right) }},\quad \rho
_{2}=\gamma _{2},\quad \rho _{3}=\gamma _{3},
\end{equation}
\begin{equation}
\omega _{1}=\frac{x^{2}}{4\left( t-t_{0}\right) }+\frac{y^{2}}{4\,t}+\frac{%
\gamma _{1}^{2}}{t_{0}}\ln \frac{t}{t-t_{0}}-\left( \gamma _{2}^{2}+\gamma
_{3}^{2}\right) \,t,
\end{equation}
\begin{equation}
\omega _{2}=\omega _{3}=\frac{\gamma _{1}^{2}}{t_{0}}\ln \frac{t}{t-t_{0}}%
-\left( \gamma _{2}^{2}+\gamma _{3}^{2}\right) \,t.
\end{equation}
This solution is partially invariant with respect to the subgroup
corresponding to the subalgebra (\ref{subSE}) with $a_{2}=a_{3}=0$ and $%
a_{1}\neq 0$, and is not reducible (unless we choose $t_{0}=0).$ For $%
t_{0}=0 $ it is invariant under rotations in the xy-plane.

\textbf{Example 5.} Let us consider again the isentropic compressible model
(formulas (\ref{IF1})--(\ref{IF2})). For the subalgebra $\frak{g}=\left\{
K_{1},K_{2},K_{3},P_{3}\right\} \,$the matrix $\Xi _{2}$ is given by 
\[
\left( 
\begin{array}{llllllll}
t & 0 & 0 & 0 & 1 & 0 & 0 & 0 \\ 
0 & t & 0 & 0 & 0 & 1 & 0 & 0 \\ 
0 & 0 & t & 0 & 0 & 0 & 1 & 0 \\ 
0 & 0 & 1 & 0 & 0 & 0 & 0 & 0
\end{array}
\right) . 
\]
Here transversality is violated also in the weak sense. The invariants of
the corresponding Lie group are $F=u_{1}t-x,G=u_{2}$ $t-y$, $a$ and $t$. The
matrix $J$ of (\ref{I.3}) is not invertible, but we can write 
\begin{equation}
u_{1}=\frac{F\left( t\right) +x}{t},\quad u_{2}=\frac{G\left( t\right) +y}{t}%
,\quad a=A\left( t\right) ,  \label{IF10}
\end{equation}
but leave $u_{3}$ general, i.e. $u_{3}=u_{3}\left( x,y,z,t\right) .$
Substituting eqs. (\ref{IF10}) into (\ref{IF1}), (\ref{IF2}) we obtain the
solution \cite{GL1} 
\begin{equation}
u_{1}=\frac{x}{t},\quad u_{2}=\frac{y}{t},\quad u_{3}=\frac{z+\lambda \left(
\xi _{1},\xi _{2}\right) }{t+t_{0}},\quad a=c\left( \frac{1}{t^{2}\left(
t+t_{0}\right) }\right) ^{\frac{1}{k}},  \label{IF11}
\end{equation}
where $\xi _{1}=\frac{x}{t},$ $\xi _{2}=\frac{y}{t}$, $\lambda $ is an
arbitrary function of $\xi _{1}$ and $\xi _{2}$, $c$ and $t_{0}$ are
constants. The rank of the matrix $Q_{a}^{\alpha }$ of the characteristics
associated to (\ref{IF11}) is equal to one, and therefore this solution is
partially invariant with respect to the subalgebra $\frak{g}$, with $\delta
=1$. Now let us check if it is reducible under any other subalgebra of the
full symmetry algebra (\ref{IFA1})--(\ref{IFA6}). To do this, it is useful
to study the kernel $K$ of the characteristic matrix $Q$ for the full
symmetry algebra (\ref{IFA1})--(\ref{IFA6}) associated to the solution (\ref
{IF11}) and to determine its generators $\left\{ \mathbf{k}_{1},...,\mathbf{k%
}_{l}\right\} $. It is clear that if at least a subspace of $K$ can be
generated by constant vectors, then the solution will be reducible with
respect to the subalgebra identified by these vectors.

In the case of the solution (\ref{IF11}), the kernel is generated by 8
vectors, each having 12 components. It is possible to show that there exists
only one constant generator, namely 
\[
\mathbf{k=}\left( 0,0,0,t_{0},0,0,0,0,0,1,0,0\right) . 
\]
This implies that the solution (\ref{IF11}) is reducible with respect to the
one--dimensional subalgebra $\left\{ K_{3}+t_{0}P_{3}\right\} $.

The concept of ''irreducibility'' of a partially invariant solutions needs
further clarification. Once a solution $u=f\left( x\right) $, partially
invariant under $G_{i}$ is found, it is of course possible to verify whether
it is invariant under some other subgroup $G_{i}^{\prime }\subset G$. Let
this be the case, let $G_{i}^{\prime }$ satisfy the strong transversality
condition, and have generic orbits of dimension $r_{i}^{\prime }$. The
standard Lie method using the subgroup $G_{i}^{\prime }\,$would then reduce
system (\ref{I.1}) to a system with $q-r_{i}^{\prime }$ variables. In
general, specially for $q-r_{i}^{\prime }>1$, we may not be able to solve
this system and the methods of partial invariance for the original subgroup $%
G_{i}\,$may be more tractable. If the invariance subgroup $G_{i}^{\prime }\,$%
does not satisfy the weak transversality condition, it may not help us at
all.

We observe that the violation of the transversality condition is \textit{not}
a necessary condition for the existence of partially invariant solutions of
a system $\Delta \left( x,u^{\left( n\right) }\right) =0$. In fact, there
could be solutions of $\Delta \,$which are not solutions of the
characteristic system , even if it is compatible as an algebraic system.

\textbf{Example 6. }As a matter of fact, a counterexample was provided by
Ondich \cite{Ondich1}, namely the two variable Laplace equation, expressed
in the following form: 
\begin{equation}
v=u_{x},\quad v_{y}=w_{x},\quad w=u_{y},\quad v_{x}=-w_{y}.  \label{LE}
\end{equation}

This system is clearly invariant with respect to the translations in the
plane, generated by the vector fields $\left\{ \partial _{x},\partial
_{y}\right\} $. The characteristic matrix $Q$ has the form 
\begin{equation}
\left( 
\begin{array}{lll}
-u_{x} & -v_{x} & -w_{x} \\ 
-u_{y} & -v_{y} & -w_{y}
\end{array}
\right) .  \label{Q1}
\end{equation}
Invariant solutions are obtained if and only if the rank of this matrix is
equal to zero. This implies that $u,v$ and $w$ are constants: 
\[
u=k,\quad v=\lambda ,\quad w=\mu ,\qquad k,\,\lambda ,\,\mu \in \Bbb{R}. 
\]

Let us impose that the rank of $Q$ is equal to 1. This means that the two
rows in eq. (\ref{Q1}) are proportional. Then solving the corresponding
equations and replacing the result into eqs. (\ref{LE}), we get the
following solution: 
\begin{equation}
\left( u,v,w\right) =\left( a\,x+b\,y+c,\,a,\,b\right) ,\qquad a,\,b\in \Bbb{%
R}.  \label{SLE}
\end{equation}
By construction, this solution is partially invariant with respect to the
group of translations in the plane, with defect $\delta =1.\,$Nevertheless,
the transversality condition is satisfied. More generally, the
transversality condition is always satisfied if we have $rank\left( \xi
_{a}^{i}\left( x,u\right) \right) =\dim $ $L.$

We mention that the previous solution (\ref{SLE}) can be also obtained
starting from the symmetry subalgebra $\left\{ \partial _{x},\partial
_{y},\partial _{u}\right\} $ for which both the weak and strong
transversality conditions are violated. In this case, the invariants are $%
I_{1}=v=a,$ $I_{2}=w=b$, where $a$ and $b$ are arbitrary constants. The
function $u$ remains arbitrary. Putting these constraints into eqs. (\ref{LE}%
), we recover immediately the solution (\ref{SLE}). However, it should be
noticed that the analysis of the rank of the characteristic matrix furnishes
a complete characterization of the invariant and partially invariant
solutions of a system of differential equations, and provides a more general
procedure than the use of the invariants. For instance, this second approach
does not allow to recognize the solution (\ref{SLE}) as a partially
invariant one with respect to the subalgebra $\left\{ \partial _{x},\partial
_{y}\right\} $.

\textbf{Example 7}. To show that a symmetry group satisfying the strong
transversality can produce both invariant and partially invariant solutions,
let us consider the Euler equations for an incompressible nonviscuous fluid
in (3+1) dimensions: 
\begin{equation}
\overrightarrow{u_{t}}+\overrightarrow{u}\cdot \nabla \overrightarrow{u}%
+\nabla p=0,  \label{Eul1}
\end{equation}
\begin{equation}
\nabla \cdot \overrightarrow{u}=0,  \label{Eul2}
\end{equation}
The symmetry group of the Euler equations is well--known \cite{Olver, Gaeta}%
. It coincides with the symmetry group of the Navier--Stokes equations,
except that it contains an additional dilation. Thus, $D$ of eq. (\ref{NS8}%
)\ is replaced by 
\begin{equation}
D_{1}=x\partial _{x}+y\partial _{y}+z\partial _{z}+t\partial _{t},
\label{D1Eul}
\end{equation}
\begin{equation}
D_{2}=t\,\partial _{t}-u_{1}\partial _{u_{1}}-u_{2}\partial
_{u_{2}}-u_{3}\partial _{u_{3}}-2p\,\partial _{p}.  \label{D2Eul}
\end{equation}
We consider here the subgroup of Galilei transformations. Its Lie algebra is
given by 
\begin{equation}
K_{1}=t\partial _{x}+\partial _{u_{1}},\quad K_{2}=t\partial _{y}+\partial
_{u_{2}},\quad K_{3}=t\partial _{z}+\partial _{u_{3}}.  \label{Gal}
\end{equation}
Here the transversality holds in the strong sense and indeed an invariant
solution will have the form 
\begin{equation}
u_{1}=\frac{x}{t}+F_{1}\left( t\right) ,\quad u_{2}=\frac{y}{t}+F_{2}\left(
t\right) ,\quad u_{3}=\frac{z}{t}+F_{3}\left( t\right) ,\quad p=P\left(
t\right) .
\end{equation}
Let us now look for partially invariant solutions of the system (\ref{Eul1}%
)--(\ref{Eul2}) with respect to the same subgroup and impose that the defect
be $\delta =2$. Writing down the characteristic system associated to (\ref
{Gal}) and imposing that $rankQ=2$, we get the following constraints on $%
u_{1}$ and $u_{2}$

\begin{equation}
u_{1}=\frac{x}{t}-\mu \,\lambda \frac{z}{t}+\mu \,\lambda u_{3}+h_{1}\left(
t\right)  \label{E1}
\end{equation}
\begin{equation}
u_{2}=\mu \,u_{3}+\frac{y}{t}-\mu \frac{z}{t}+h_{2}\left( t\right)
\label{E2}
\end{equation}
but $u_{3}=u_{3}\left( x,y,z,t\right) $ and $p=p\left( x,y,z,t\right) $
remain arbitrary.

Substituting the relations (\ref{E1})--(\ref{E2}) into the Euler equations,
and choosing for simplicity $h_{1}=h_{2}=0,\,$we obtain the solution: 
\begin{eqnarray}
u_{1} &=&\frac{1}{t\,\left[ \mu ^{2}\left( 1+\lambda ^{2}\right) +1\right] }%
\left\{ x\,\left[ \mu ^{2}\left( 1-2\lambda ^{2}\right) +1\right] -3\lambda
\mu \,\left( \mu y+z\right) \right\} +  \nonumber \\
&&\lambda \,\mu \,t^{2}F\left( \frac{\lambda y-x}{t},\frac{\lambda \,\mu
\,z-x}{t}\right) ,  \label{SE1}
\end{eqnarray}
\begin{eqnarray}
u_{2} &=&\frac{1}{t\,\left[ \mu ^{2}\left( 1+\lambda ^{2}\right) +1\right] }%
\left\{ y\,\left[ \mu ^{2}\left( \lambda ^{2}-2\right) +1\right] -3\mu
\,\left( \lambda \mu x+z\right) \right\} +  \nonumber \\
&&\,\mu \,t^{2}F\left( \frac{\lambda y-x}{t},\frac{\lambda \,\mu \,z-x}{t}%
\right) ,  \label{SE2}
\end{eqnarray}
\begin{eqnarray}
u_{3} &=&\frac{1}{t\,\left[ \mu ^{2}\left( 1+\lambda ^{2}\right) +1\right] }%
\left\{ z\,\left[ \mu ^{2}\left( 1+\lambda ^{2}\right) -2\right] -3\left(
\lambda x+y\right) \right\} +  \nonumber \\
&&\,t^{2}F\left( \frac{\lambda y-x}{t},\frac{\lambda \,\mu \,z-x}{t}\right) ,
\label{SE3}
\end{eqnarray}
\begin{equation}
p=-\frac{3\mu ^{2}}{t^{2}\left( \lambda ^{2}\mu ^{2}+\mu ^{2}+1\right) \,}%
\left( \lambda x+y+\frac{z}{\mu }\right) ^{2}+p\left( t\right) ,  \label{SE4}
\end{equation}
where $\lambda ,\mu $ are constants, $F$ is an arbitrary function of its
arguments and $p$ is an arbitrary function of $t$. We have checked
explicitly that, if $F$ is kept arbitrary, this solution is not invariant
under any subgroup of the symmetry group. Thus, it represents an irreducible
partially invariant solution of the Euler equations of defect $\delta =2,$
with respect to a subgroup satisfying the strong transversality condition.
If 
\[
F=\left( \frac{\lambda y-x}{t}\right) ^{\frac{3a+2b}{a}}\Phi \left( \frac{%
\lambda y-x}{\lambda \,\mu \,z-x}\right) ,\qquad a,b\in \Bbb{R}, 
\]
where $\Phi \,$is an arbitrary function of its argument, then the solution (%
\ref{SE1})--(\ref{SE4}) is invariant under the subgroup generated by $%
aD_{1}+bD_{2},\,$where $D_{1}$ and $D_{2}$ are defined in eqs. (\ref{D1Eul}%
)--(\ref{D2Eul}).

However, this subgroup provides a reduced system with three independent
variables that would be very difficult to solve.

Particularly interesting is the case when partially invariant solutions can
be found that satisfy weak but not strong transversality. Indeed, imposing
weak transversality basically means that a class of functions $u=f\left(
x\right) $ is chosen in such a way that the characteristic system is
algebraically compatible. This condition is of course not sufficient to
guarantee the invariance of these functions under the action of the
considered group $G$. Indeed, if we compute the rank of the matrix $%
Q_{a}^{\alpha }$ on this class of functions, in general it will be not equal
to zero. Then, once weak transversality is satisfied, we can choose either
to have group invariant solutions, using the method outlined in Section 3,
or to use the class of functions $u=f\left( x\right) $ to get partially
invariant solutions.

In the next two examples, we will see how this approach can be used to
obtain in a simple and straightforward way new classes of solutions of
hydrodynamic systems.

\textbf{Example 8}. In Example 2 we studied the algebra (\ref{g2}) which is
a subalgebra of the symmetry algebra of both the Euler and the
Navier--Stokes equations. We showed that the requirement of weak
transversality implies 
\begin{equation}
u_{1}=k\frac{x}{t},\qquad u_{2}=k\frac{y}{t},  \label{E81}
\end{equation}
\begin{equation}
u_{3}=u_{3}\left( x,y,z,t\right) ,\qquad p=p(x,y,z,t).  \label{E82}
\end{equation}
These formulas define a class of functions which is partially invariant with
defect $\delta =2$ with respect to the subalgebra (\ref{g2}). At this stage,
we can choose either to have group invariant solutions or partially
invariant ones. Indeed, in Example 2 we forced the class of functions (\ref
{E81})--(\ref{E82}) to be a solution of the characteristic system, and then
substituting the obtained expressions (\ref{S21})--(\ref{S24}) into the
system (\ref{NS1})--(\ref{NS2}) we constructed the group invariant solutions
(\ref{S25})--(\ref{S26}).

Another possibility is to substitute formulas (\ref{E81})--(\ref{E82})
directly into the Euler equations (\ref{Eul1})--(\ref{Eul2}) or the
Navier--Stokes equations (\ref{NS1})--(\ref{NS2}) without requiring further
invariance properties. For the case of the Euler equations we get the
constraints 
\begin{equation}
t^{2}p_{x}+k\,(k-1)\,x=0,  \label{E83}
\end{equation}
\begin{equation}
t^{2}p_{y}+k\,(k-1)\,y=0,
\end{equation}
\begin{equation}
u_{3z}+\frac{2k}{t}=0,
\end{equation}
\begin{equation}
u_{3t}+u_{3}u_{3z}+\frac{k}{t}\left( x\,u_{3x}+y\,u_{3y}\right) +p_{z}=0.
\label{E86}
\end{equation}
We solve this system and obtain the following partially invariant solution
of the Euler equations: 
\begin{equation}
u_{1}=k\frac{x}{t},\qquad u_{2}=k\frac{y}{t},
\end{equation}
\begin{equation}
u_{3}=-\frac{2kz}{t}+x^{2}F\,\left( tx^{-\frac{1}{k}},\frac{y}{x}\right)
\end{equation}
\begin{equation}
p=-\frac{k\,\left( k-1\right) \,\left( x^{2}+y^{2}\right) }{2\,t^{2}}-\frac{%
k\,\left( 2k+1\right) \,z^{2}}{t^{2}}+f\,\left( t\right) ,
\end{equation}
where $\xi =tx^{-\frac{1}{k}},\eta =\frac{y}{x}$ and $f\left( t\right) $ and 
$F\,\left( \xi ,\eta \right) $ are arbitrary functions of their arguments.
This solution is irreducible for a generic function $F$. If $F$ satisfies
the equation 
\[
\left[ \left( c_{1}\frac{k-1}{k}+c_{2}\right) \xi -c_{3}\frac{\xi \eta }{k}-%
\frac{c_{4}}{k}\xi ^{k+1}+c_{5}\xi ^{k}\right] F_{\xi }+ 
\]
\[
\left[ -c_{3}\left( \eta ^{2}+1\right) -c_{4}\xi ^{k}\eta +c_{5}\xi
^{k}\right] F_{\eta }+\left( 2c_{1}+c_{2}+2c_{3}\eta \right) F-c_{6}\left(
4k+1\right) \xi ^{2k}=0, 
\]
then it is invariant under the subgroup generated by

\[
X=c_{1}D_{1}+c_{2}D_{2}+c_{3}L_{3}+c_{4}B_{1}+c_{5}B_{2}+c_{3}B_{3}, 
\]
where $c_{1},...,c_{6}$ are real constants, $D_{1}$ and $D_{2}$ are the
dilations given by eqs. (\ref{D1Eul})--(\ref{D2Eul}), $L_{3}$ is the
generator (\ref{NS11}), and the functions appearing in the boosts (\ref{NS3}%
)--(\ref{NS5}) are now monomials in $t$, namely 
\[
\alpha =t^{k},\quad \beta =t^{k},\quad \gamma =t^{2k+1}. 
\]

The same procedure can be applied to the Navier--Stokes equations. Repeating
the previous steps, we obtain the following solution: 
\begin{equation}
u_{1}=k\frac{x}{t},\qquad u_{2}=k\frac{y}{t},
\end{equation}
\begin{equation}
u_{3}=-\frac{2\,k\,z}{t}+\alpha \left( x,y,t\right)
\end{equation}
\begin{equation}
p=-\frac{k\,\left( k-1\right) \,\left( x^{2}+y^{2}\right) }{2\,t^{2}}-\frac{%
k\,\left( 2k+1\right) \,z^{2}}{t^{2}}+f\,\left( t\right) ,
\end{equation}
where $\alpha \left( x,y,t\right) $ satisfies the following equation 
\begin{equation}
\alpha _{t}+\frac{k}{t}\left( x\,\alpha _{x}+y\,\alpha _{y}-2\alpha \right)
-\nu \,\left( \alpha _{xx}+\alpha _{yy}\right) =0.  \label{LNS}
\end{equation}
We thus obtain a large class of partially invariant solutions of the
Navier--Stokes equations parametrized by the solutions of the \textit{linear}
partial differential equation (\ref{LNS}).

\textbf{Example 9. }Partially invariant solutions with weak transversality
can be also found for the case of the compressible fluid model (\ref{IF1})--(%
\ref{IF2}). Let us again consider the subalgebra $\left\{
L_{3},F+G,K_{1},K_{2}\right\} $ and the corresponding weak transversality
condition (\ref{IF4}). Substituting into equations (\ref{IF1})--(\ref{IF2})
we obtain a coupled system of quasilinear first order PDE's, namely: 
\[
a_{x}=0,\qquad a_{y}=0, 
\]
\[
u_{3t}+u_{3}\,u_{3z}+\frac{x}{t}u_{3x}+\frac{y}{t}u_{3y}+k\,a\,a_{z}=0 
\]
\begin{equation}
a_{t}+u_{3}\,a_{z}+\frac{a}{k}\left( \frac{2}{t}+u_{3z}\right) =0.
\label{IF12}
\end{equation}

In particular, if we assume $a_{z}=0$, we reobtain the solution (\ref{IF11}%
). However, the system (\ref{IF12}) allows much more general solutions.

\section{Conclusions}

The main conclusion of this article is that one can do considerably more
with the symmetry group $G$ of a system of partial differential equations
than apply the standard method for finding group invariant solutions.

Indeed, let us assume that the largest group $G$ (of local Lie point
transformations) leaving the system (\ref{I.1}) invariant has been found and
its subgroups classified. For each subgroup $G_{0}$, or its Lie algebra $%
L_{0}$ we should proceed as follows.

1. Obtain invariant solutions. First check whether the transversality
condition (\ref{10}) is satisfied (in the strong sense). If it is, we apply
Lie's classical method. This is always possible since transversality assures
that the rank condition (\ref{I.3}) is satisfied. If the (strong)
transversality condition is not satisfied, we may still be able to obtain
solutions invariant under $G_{0}$ \cite{AFT}, by imposing ''weak
transversality'' on solutions, as described in Section 2 above. This is
illustrated in Section 3 by Examples 1,2 and 3.

2. Obtain partially invariant solutions. These can be obtained by (at least)
three complementary methods. If the transversality condition (\ref{10}) is
not satisfied and weak transversality cannot be imposed, then the
characteristic system (\ref{I.4}) is not consistent and the rank condition (%
\ref{I.3}) for the invariants is not satisfied. We then choose a subset of $%
q^{\prime }<q$ invariants such that we can express $q^{\prime }$ dependent
variables in terms of invariants. The remaining $q-q^{\prime }$ variables $%
u_{\alpha }$ are considered as functions of all the original variables $%
x_{1},...,x_{p}$. Examples 4 and 5 of Section 4 are of this type, as are
those of Ref. \cite{MW}, \cite{MSW}.

If the transversality condition is satisfied for $G_{0}$ we may still be
able to obtain partially invariant solutions, in addition to the invariant
ones. Instead of imposing that the matrix of characteristics (\ref{I.4})
have rank zero (i.e. that all equations (\ref{I.4}) be satisfied) we require
that on solutions we have $rankQ=\delta $, with $\delta =1,2,...,$ as the
case may be. This rank condition must be solved explicitly for $u_{\alpha }$
and the result substituted into eq. (\ref{I.1}) (see Examples 6,7). The
third possibility is to impose weak transversality (if possible), but still
to impose $rankQ=\delta \geq 1$ (see Examples 8 and 9).

3. Go beyond invariant and partially invariant solutions, either by the
method of group foliation \cite{Ovs, NS, MShW}, or by other methods not
using the symmetry group $G$ \cite{Clark-Krus}--\cite{Sacco}.

Missing at this stage are clear criteria that tell us which approach will be
fruitful. Furthermore, the same solution may be obtained by different
methods and it is not clear which of these method will lead to the least
amount of calculations. For instance, solutions partially invariant under
some subgroup $G_{1}\subset G$ have been called ''reducible'' \cite{MW}--%
\cite{Ondich2}, \cite{GL1}--\cite{GL3} if they are actually invariant under
some other subgroup $G_{2}\subset G$. However, it may be more difficult to
use $G_{2},$ than $G_{1}$, specially if the dimension of $G_{2}$ is small
with respect to the number of independent variables.

\textbf{Acknowledgements}

One of the authors (P. W.) thanks Professors Ian Anderson and Mark Fels for
interesting discussions. The research of A. M. Grundland and of P.
Winternitz was partially supported by NSERC of Canada and FCAR\ du
Qu\'{e}bec.


\begin{thebibliography}{99}
\bibitem{Olver}  P. J. Olver, \textit{Application of Lie groups to
Differential Equations} (Springer--Verlag, New York, 1986).

\bibitem{Bluman}  G. W. Bluman and S. Kumei, \textit{Symmetries and
Differential Equations} (Springer--Verlag, New York, 1989).

\bibitem{Ovs}  L. V. Ovsiannikov, \textit{Group Properties of Differential
Equations }(in russian), (Novosibirsk, 1962); \textit{Group Analysis of
Differential Equations }(Academic Press, 1982).

\bibitem{Ibrag}  N. H. Ibragimov, \textit{Transformation Groups Applied to
Mathematical Physics}, (D. Reidel Publishing Company, 1985).

\bibitem{Gaeta}  G. Gaeta, \textit{Nonlinear Symmetries and Nonlinear
Equations} (Kluwer, Dordrecht, 1994).

\bibitem{Wintern}  P. Winternitz, in \textit{Partially integrable Evolution
Equations in Physics}, pp. 515--567, edited by R. Conte and N. Boccara
(Kluwer, Amsterdam, 1990).

\bibitem{Wintern2}  P. Winternitz, Lie Groups and Solutions of Nonlinear
Partial Differential Equations, in \textit{Integrable Systems, Quantum
Groups and Quantum Field Theory}, pp. 429--495, edited by A. Ibort and M. A.
Rodriguez (Kluwer, Dordrecht, 1993).

\bibitem{Vino}  I. S. Krasil'shchik and A. M. Vinogradov (edts.), \textit{%
Symmetries and Conservation Laws for Differential Equations of Mathematical
Physics,} (American Mathematical Society, 1999).

\bibitem{SSY}  A. F. Sidorov, V. P. Shapeev and N. N. Yanenko, \textit{The
Method of Differential Constraints and its Applications in Gas Dynamics} (in
russian), (Nauka, Novosibirsk, 1984).

\bibitem{Olver-Ros1}  P. J. Olver and P. Rosenau, The construction of
special solutions to partial differential equations, Phys. Lett. A \textbf{%
114}, 107 (1986).

\bibitem{Olver-Ros2}  P. J. Olver and P. Rosenau, Group--invariant solutions
of differential equations, SIAM\ J. Appl. Math. \textbf{47}, 263 (1987).

\bibitem{AFT}  I. M. Anderson, M. E. Fels and C. G. Torre, Group invariant
solutions without transversality, Comm. Math. Phys. \textbf{212}, 653 (2000).

\bibitem{Ovs2}  L. V. Ovsiannikov, Partial invariance, Sov. Math. Dokl. 
\textbf{10}, 780 (1969).

\bibitem{MW}  L. Martina and P. Winternitz, Partially invariant solutions of
nonlinear Klein--Gordon and Laplace equations, J. Math. Phys. \textbf{33},
2718 (1992).

\bibitem{MSW}  L. Martina, G. Soliani and P. Winternitz, Partially invariant
solutions of a class of nonlinear Schr\"{o}dinger equations, J. Phys. A:
Math. Gen. \textbf{25}, 4425 (1992).

\bibitem{GL1}  A. M. Grundland and L. Lalague, Invariant and
partially--invariant solutions of the equations describing a non--stationary
and isentropic flow for and ideal and compressible fluid in (3+1)
dimensions, J. Phys. A: Math. Gen \textbf{29}, 1723 (1996).

\bibitem{GL2}  A. M. Grundland and L. Lalague, Lie subgroups of symmetry
groups of fluid dynamics and magnetohydrodynamics equations, Can. J. Phys. 
\textbf{73}, 463 (1995).

\bibitem{GL3}  A. M. Grundland and L. Lalague, Lie subgroups of the symmetry
group of the equations describing a nonstationary and isentropic flow:
invariant and partially invariant solutions, Can. J. Phys. \textbf{72}, 362
(1994).

\bibitem{Ondich1}  J. Ondich, The reducibility of partially invariant
solutions of systems of partial differential equations, Euro. J. Appl. Math. 
\textbf{6}, 329 (1995).

\bibitem{Ondich2}  J. Ondich, A differential constraint approach to partial
invariance, Euro. J. Appl. Math. \textbf{6}, 631 (1995).

\bibitem{Ibrag2}  N. H. Ibragimov, Generalized motions in Riemann spaces,
Soviet Math. Dokl. \textbf{9}, 21 (1968).

\bibitem{Ibrag3}  N. H. Ibragimov, Groups of generalized motions, Soviet
Math. Dokl. \textbf{10}, 538 (1969).

\bibitem{SDR}  C. C. A. Sastri, K. A. Dunn and D. R. K. S. Rao,
Ovsiannikov's method and the construction of partially invariant solutions,
J. Math. Phys. \textbf{28}, 1473 (1987).

\bibitem{Clark-Krus}  P. Clarkson and M. Kruskal, New similarity solutions
of the Boussinesq equation J. Math. Phys. \textbf{30}, 2201 (1989).

\bibitem{CM}  P. Clarkson and E. Mansfield, Algorithms for the nonclassical
method of symmetry reductions, SIAM\ J. Appl. Math \textbf{54}, 1693 (1994).

\bibitem{CW}  P. Clarkson and P. Winternitz, Symmetry Reduction and Exact
Solutions of Nonlinear Partial Differential Equations, in \textit{The
Painlev\'{e} Property. One Century Later}, R. Conte ed. (Springer, New York,
1999).

\bibitem{Pucci}  E. Pucci, Similarity reductions of partial differential
equations, J. Phys. A \textbf{25}, 2631 (1992).

\bibitem{Levi-Wint}  D. Levi and P. Winternitz, Nonclassical symmetry
reduction:example of the Boussinesq equation, J. Phys. A \textbf{22}, 2915
(1989).

\bibitem{Blum-Cole}  G. W. Bluman and J. D. Cole, The general similarity of
the heat equation, J. Math. Mech. \textbf{18}, 1025 (1969).

\bibitem{ABH}  D. J. Arrigo, P. Broadbridge and J. M. Hill, Nonclassical
symmetry solutions and the methods of Bluman--Cole and Clarkson--Kruskal, J.
Math. Phys \textbf{34}, 4692 (1993).

\bibitem{Sacco}  G. Saccomandi, Potential Symmetries and direct reduction
methods of order two, J. Phys. A: Math. Gen. \textbf{30}, 2211 (1997).

\bibitem{NS}  Yu. Nutku and M. B. Sheftel, Differential invariants and group
foliation for the complex Monge--Ampere equation, J. Phys A \textbf{34}, 137
(2001).

\bibitem{MShW}  L. Martina, M. B. Sheftel and P. Winternitz, Group foliation
and non--invariant solutions of the heavenly equation, J. Phys A \textbf{34}%
, 9243 (2001).

\bibitem{CG}  G. Cicogna and G. Gaeta, Partial Lie--point symmetries of
differential equations, J. Phys. A \textbf{34}, 491 (2001).

\bibitem{AC}  M. J. Ablowitz and P. A. Clarkson, \textit{Solitons, Nonlinear
Evolution Equations and Inverse Scattering} (Cambridge University Press,
Cambridge, 1991).

\bibitem{BRK}  G.W. Bluman, G.J. Reid and S. Kumei, New classes of
symmetries for partial differential equations,\ J. Math. Phys.\textbf{\ 29},
806 (1988).

\bibitem{KV}  I.S. Krasil'shchik and\ A.M. Vinogradov, Nonlocal trends in
the geometry of differential equations: Symmetries, conservation laws and
B\"{a}cklund transformations,\ Acta Appl. Math. \textbf{15}, 161 (1989).

\bibitem{LLST}  M. Leo, R.A. Leo, G. Soliani and P. Tempesta, On the
Relation between Lie Symmetries and Prolongation Structures of Nonlinear
Field Equations: Non-Local Symmetries, Prog. Theor. Phys. \textbf{105}, 77
(2001).

\bibitem{GLST}  V. Grassi, R.A. Leo, G. Soliani and P. Tempesta, Vortices
and invariant surfaces generated by symmetries for the 3D Navier--Stokes
equations, Physica A \textbf{286}, 79 (2000).

\bibitem{Ll}  S. P. Lloyd, The infinitesimal group of the Navier--Stokes
equations, Acta Mech. \textbf{38}, 85 (1981).

\bibitem{Ovs3}  L. V. Ovsiannikov, \textit{Lectures on Principles of Gas
Dynamics} (Nauka, Moskow, 1981).

\bibitem{Ince}  E. L. Ince, \textit{Ordinary Differential Equations} (Dover,
New York, 1956).

\bibitem{Olver2}  P. J. Olver, Symmetry and Explicit Solutions of Partial
Differential Equations, Appl. Num. Math. \textbf{10}, 307 (1992).

\bibitem{Lang}  S. Lang, \textit{Linear Algebra (}Springer--Verlag, New
York, 1987).

\bibitem{BZ}  A. L. Berkhoer and V. E. Zakharov, Self excitation of waves
with different polarizations in nonlinear media, Sov. Phys. JETP, \textbf{31}%
, 486 (1970).

\bibitem{Benney}  D. J. Benney and G. J. Roskes, Wave instabilities, Stud.
Appl. Math. \textbf{48}, 377 (1969).

\bibitem{Grimshaw}  R. Grimshaw, Slowly varying solitary waves. II Nonlinear
Schr\"{o}dinger equation, Proc. R. Soc. London, Ser. A, \textbf{368}, 377
(1979).

\bibitem{Phillips}  O. M. Phillips, \textit{The Dynamics of the Upper Ocean}
(Cambridge University Press, London, 1977).

\bibitem{Rem}  M. Remoissenet, \textit{Waves Called Solitons. Concepts and
Experiments} (Springer, Berlin, 1999).

\bibitem{SW}  A. Sciarrino and P. Winternitz, Symmetries and solutions of
the vector nonlinear Schr\"{o}dinger equation, Il Nuovo Cimento \textbf{112}
B, 853 (1997).
\end{thebibliography}
\end{document}